\documentclass[twocolumn,superscriptaddress,floatfix,preprintnumbers,prd,nofootinbib,hyperref]{revtex4-1}

\pdfoutput=1

\usepackage{graphicx,color}
\usepackage{amsmath,amssymb,eqnarray}
\usepackage{hyperref} 

\hypersetup{
    colorlinks=true,
    citecolor=[rgb]{.1, .7, .6},
    linkcolor=[rgb]{.2, .55, .95},
    filecolor=magenta,
    urlcolor=[rgb]{.1, .7, .6},
}
\newcommand{\unit}[1]{\,\mathrm{#1}}
\newcommand{\me}{\mathrm{e}}

\renewcommand{\vec}[1]{\boldsymbol{#1}}
\def\epsilon{\varepsilon}
\def\theta{\vartheta}

\newcommand{\ie}{{\it i.e.~}}  \newcommand{\eg}{{\it e.g.~}}
\newcommand{\cf}{{\it cf.~}}

\begin{document}

\title{Adiabatic Axion-Photon Mixing Near Neutron Stars}

\author{Jonas Tjemsland}
\affiliation{Department of Physics, Norwegian University of Science and Technology, Høgskoleringen 5, 7491, Trondheim, Norway}

\author{Jamie McDonald}
\affiliation{Centre for Cosmology, Particle Physics and Phenomenology
(CP3), Universit \'{e} Catholique de Louvain, Chemin du cyclotron 2,
Louvain-la-Neuve B-1348, Belgium}
\affiliation{Department of Physics and Astronomy, University of Manchester, Manchester M13 9PL, UK}

\author{Samuel J. Witte}
\affiliation{Rudolf Peierls Centre for Theoretical Physics, University of Oxford, Parks Road, Oxford OX1 3PU, UK}
\affiliation{Departament de F\'{i}sica Qu\`{a}ntica i Astrof\'{i}sica and Institut de Ciencies del Cosmos (ICCUB) ,
Universitat de Barcelona, Diagonal 647, E-08028 Barcelona, Spain}


\begin{abstract}
      One of the promising new proposals to search for axions in astrophysical environments is to look for narrow radio lines produced from the resonant conversion of axion dark matter falling through the magnetospheres of neutron stars. For sufficiently strong magnetic fields, axion masses in the $\mathcal{O}(10\mu{\rm eV)}$ range, and axion-photon couplings $g_{a\gamma} \gtrsim 10^{-12} \, {\rm GeV^{-1}}$, the conversion can become hyper-efficient, allowing axion-photon and photon-axion transitions to occur with $\mathcal{O}(1)$ probabilities. Despite the strong mixing between these particles, the observable radio flux emanating from the magnetosphere is expected to be heavily suppressed -- this is a consequence of the fact that photons sourced by infalling axions have a high probability of converting back into axions before escaping the magnetosphere. In this work, we study the evolution of the axion and photon phase space near the surface of highly magnetized neutron stars in the adiabatic regime, quantifying for the first time the properties of the radio flux that arise at high axion-photon couplings. We show that previous attempts to mimic the scaling in this regime have been overly conservative in their treatment, and that the suppression can be largely circumvented for radio observations targeting neutron star populations.
\end{abstract}

\maketitle

\section{Introduction}

Axions and axion-like-particles are amongst the most compelling candidates for new fundamental physics; this is because these particles provide a simple solution to the strong-CP problem \cite{Peccei:1977ur,Kim:1979if,Shifman:1979if,Dine:1981rt,Zhitnitsky:1980tq}, an explanation for dark matter (via the misalignment mechanism \cite{Dine:1982ah,Abbott:1982af,Preskill:1982cy} or the decays of topological defects~\cite{Preskill:1982cy,Abbott:1982af,Dine:1982ah,Hiramatsu:2012gg,Kawasaki:2014sqa,Klaer:2017ond,Gorghetto:2020qws,Buschmann:2021sdq}), and appear abundantly in well-motivated high-energy extensions of the Standard Model, such as String Theory~\cite{Arvanitaki:2009fg,Witten:1984dg,Cicoli:2012sz,Conlon:2006tq,Svrcek:2006yi}. 

There are growing experimental efforts across the globe to search for dark matter axions using haloscopes~\cite{Sikivie:1983ip}, which typically attempt to measure the coupling of axions to photons, given by $\mathcal{L}_{a \gamma} = -\frac{1}{4} g_{a \gamma} a F_{\mu \nu}\tilde{F}^{\mu \nu}$, where $F_{\mu \nu}$ is the photon field strength tensor, $a$ is the axion field, and $g_{a \gamma}$ is a dimensionful coupling constant. The most successful approach to date involves constructing a small cavity whose electromagnetic modes can be tuned to match the frequency of the background axion field~\cite{DePanfilis:1987dk,Wuensch:1989sa,Hagmann:1990tj,ADMX:2009iij,ADMX:2018gho,ADMX:2009iij,ADMX:2018gho,ADMX:2019uok,ADMX:2021nhd,ADMX:2018ogs,ADMX:2021mio,Crisosto:2019fcj,Lee:2020cfj,Jeong:2020cwz,Lee:2022mnc,Kim:2022hmg,Brubaker:2017rna,HAYSTAC:2018rwy,HAYSTAC:2020kwv,Alesini:2019ajt,Alesini:2020vny,McAllister:2017lkb,Quiskamp:2022pks,CAST:2020rlf,TASEH:2022vvu,Grenet:2021vbb}, however a variety of alternative ideas have also emerged which attempt to overcome the challenges of conventional cavity searches, allowing laboratory experiments to probe a broader range of axion masses and interactions (see \eg \cite{ALPHA:2022rxj,BREAD:2021tpx,Beurthey:2020yuq,DMRadio:2022pkf,Marsh:2022fmo,McDonald:2021hus,Schutte-Engel:2021bqm}). 
Another approach is to look for signatures of axions in astrophysical environments (see \eg\cite{Boddy:2022knd,Adams:2022pbo,Baryakhtar:2022hbu} for recent reviews); these techniques are highly complementary to laboratory experiments since they are often capable of probing a wider range of axion masses, rely on different assumptions of the underlying distribution of axion dark matter, and can be used to break intrinsic degeneracies that arise in terrestrial searches. 

Amongst the more promising indirect axion searches proposed in recent years is the idea of looking for radio signatures that arise as axions pass through the magnetospheres of neutron stars. Here, the large magnetic fields and dense ambient plasma can dramatically amplify the interactions between axions and photons, giving rise to a variety of distinctive features, including narrow radio spectral lines~\cite{Pshirkov:2007st,Hook:2018iia,Huang:2018lxq,Leroy:2019ghm,Safdi:2018oeu,Battye:2019aco,Foster:2020pgt,Prabhu:2020yif,Foster:2022fxn,Witte:2021arp,Millar:2021gzs,Battye:2021yue,Battye:2023oac,Xue:2023ejt,Caputo:2023cpv}, an excess of broadband radio emission~\cite{Prabhu:2021zve,Noordhuis:2022ljw,Noordhuis:2023wid}, and radio transients~\cite{Buckley:2020fmh,Edwards:2020afl,Bai:2021nrs,Nurmi:2021xds,Witte:2022cjj,Prabhu:2023cgb}. Recent observational efforts searching for some of these signatures have already been used to set highly competitive constraints on the axion-photon coupling (see \eg\cite{Foster:2022fxn,Noordhuis:2022ljw,Battye:2023oac}).

Axions are generally thought of as feebly-interacting particles, implying that their interactions are, in most contexts, only expected to  induce small perturbative effects on the systems of interest. For example, axion dark matter falling through a neutron star magnetosphere is typically expected to pass through the entire system unperturbed, only on rare occasions sourcing low-energy radio photons.
Despite being a rare process, however, this signal can shine through astrophysical backgrounds thanks to: (i) the distinctive spectral shape of the radio signal, manifested as an extremely narrow spectral line (which sharply contrasts against smooth astrophysical backgrounds), and (ii) a large local axion number density---potentially exceeding $\sim 10^{20} \, {\rm cm^{-3}}$ ---which can compensate for the inefficiency of axion-photon conversion.

Early observational campaigns~\cite{Darling:2020uyo,Darling:2020plz,Foster:2020pgt,Battye:2021yue} looking for radio lines produced from axion dark matter derived limits on the axion-photon coupling in this `perturbative limit', \ie they worked under the assumption that the axion-photon conversion probability was always small, $P_{a\rightarrow \gamma} \ll 1$, implying that the radio luminosity scales as $L_{\rm radio} \propto g_{a\gamma}^2$. It was only recently pointed out in Ref.~\cite{Foster:2022fxn} that these assumptions can be strongly violated, particularly at large (but still viable) axion-photon couplings and for pulsars with strong magnetic fields. Instead of occasionally sourcing an on-shell photon, axions falling through the magnetosphere are expected to convert with $\mathcal{O}(1)$ probability. The story doesn't end there, however, as the newly produced photons will themselves encounter resonances\footnote{Note that the non-resonant axion-photon conversion is heavily suppressed in these systems and thus can be fully neglected.}, converting back to axions with $\mathcal{O}(1)$ probability.
In the large $g_{a\gamma}$ limit, the expectation is that photons typically convert back into axions before escaping the magnetosphere, resulting in highly suppressed radio luminosity.

As a first attempt to include these `adiabatic conversions' into the calculation of the radio flux, Ref.~\cite{Foster:2022fxn} adopted the simplifying assumption that photon production at each resonance could be approximated by using a net effective conversion probability which is set by the product of the survival probability with the conversion probability, \ie $P^{\rm eff}_{a\rightarrow \gamma} = (1 - P_{a \rightarrow \gamma}) \times  P_{a \rightarrow \gamma}$. This approximation, which leads to an exponential suppression (see Eq.~\eqref{eq:LZ}) in the large coupling limit, is strictly speaking only valid when the axion-photon resonances take place on a spherical surface centered about the neutron star, and when the conversion probability is equivalent for all axions and all photons, \ie it depends only on the radial distance from the neutron star. For realistic systems, neither of these assumptions hold, and it remains unclear how well this approximation reflects the true rate of photon production in the adiabatic regime.

In this manuscript, we develop an algorithm capable of carefully tracking the evolution of the axion and photon phase space around neutron stars, and characterize, for the first time, the scaling and properties of the radio flux produced from adiabatic resonant conversion of axion dark matter in the magnetospheres of neutron stars. For large axion-photon couplings and small axion masses, our algorithm recovers the approximate exponential suppression predicted in~\cite{Foster:2022fxn}. This suppression, however, is only valid for a range of couplings -- instead of being exponentially suppressed  in the limit that $g_{a\gamma} \rightarrow \infty$,  the radio flux instead asymptotes to a fixed finite value.

That being said, the suppression of the radio flux can be partially avoided at larger axion masses, where there is more room for axions to traverse the magnetosphere in such a way that they encounter only one level crossing (see top row of Fig.~\ref{fig:examples}). These axions contribute to the radio flux, but are limited in number, implying the radio luminosity is phase-space suppressed relative to the naive perturbative approach in which re-conversions are neglected. 
We show that the observed suppression is crucially dependent on the geometry of the resonant surface around the neutron star, and provide approximate expressions which can be used to extrapolate the functional scaling of the radio luminosity from the non-adiabatic to the adiabatic regime, thereby evading the need for complex numerical analyses in the high coupling limit. This represents an important step in solidifying the limits derived in~\cite{Foster:2022fxn}, and in establishing techniques that allow for future radio surveys to probe axions at large couplings.

The structure of this paper is as follows. In Sec.~\ref{sec:RayTracing} we provide a general overview of mixing and propagation of axions and photons in magnetized plasmas. In Sec.~\ref{sec:Trees} we discuss the algorithm that we develop to  self-consistently track the worldlines and production probabilities of particles sourced by infalling axions. We implement these techniques in Sec.~\ref{sec:Results} and examine the behavior and scaling of the radio flux for adiabatic axion-photon conversion. In Sec.~\ref{sec:conclusions} we give our conclusions. 

\section{Axion-Photon mixing around neutron stars}\label{sec:RayTracing}

\begin{figure*}
        \includegraphics[width=.49\textwidth]{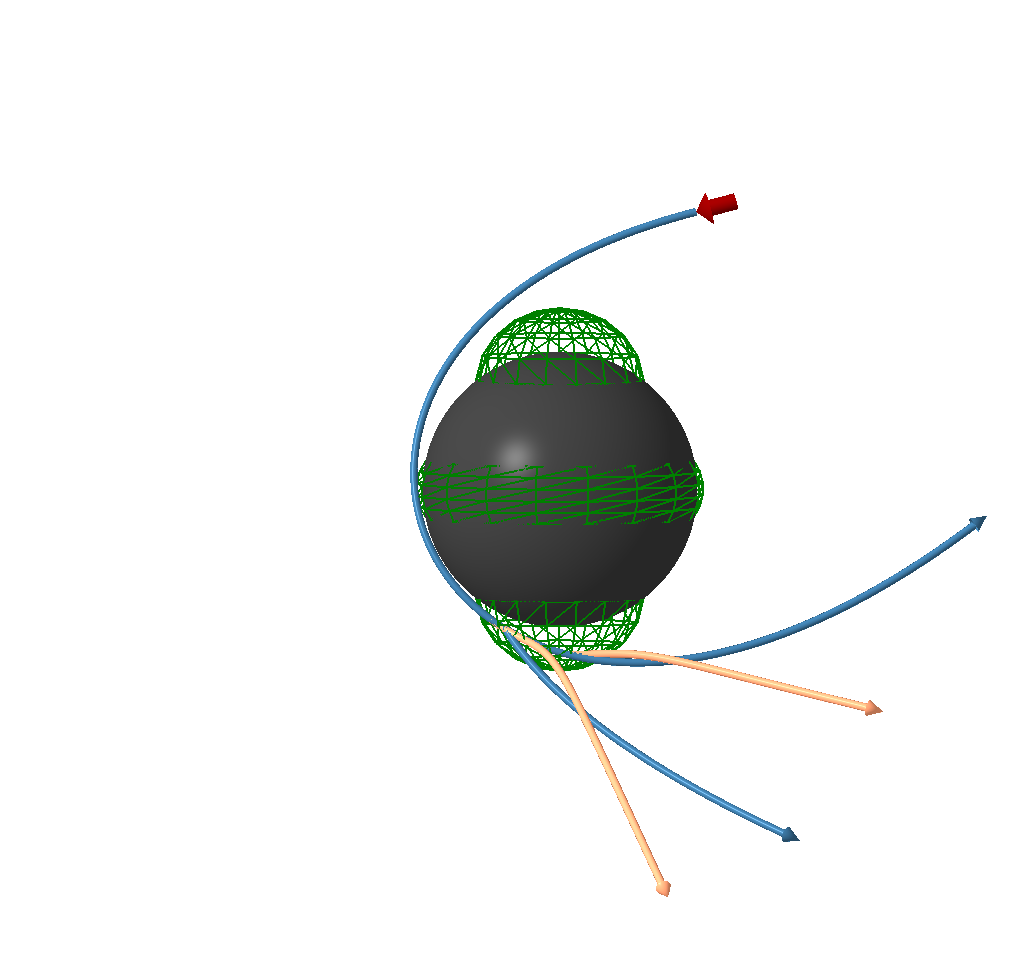}
        \includegraphics[width=.49\textwidth]{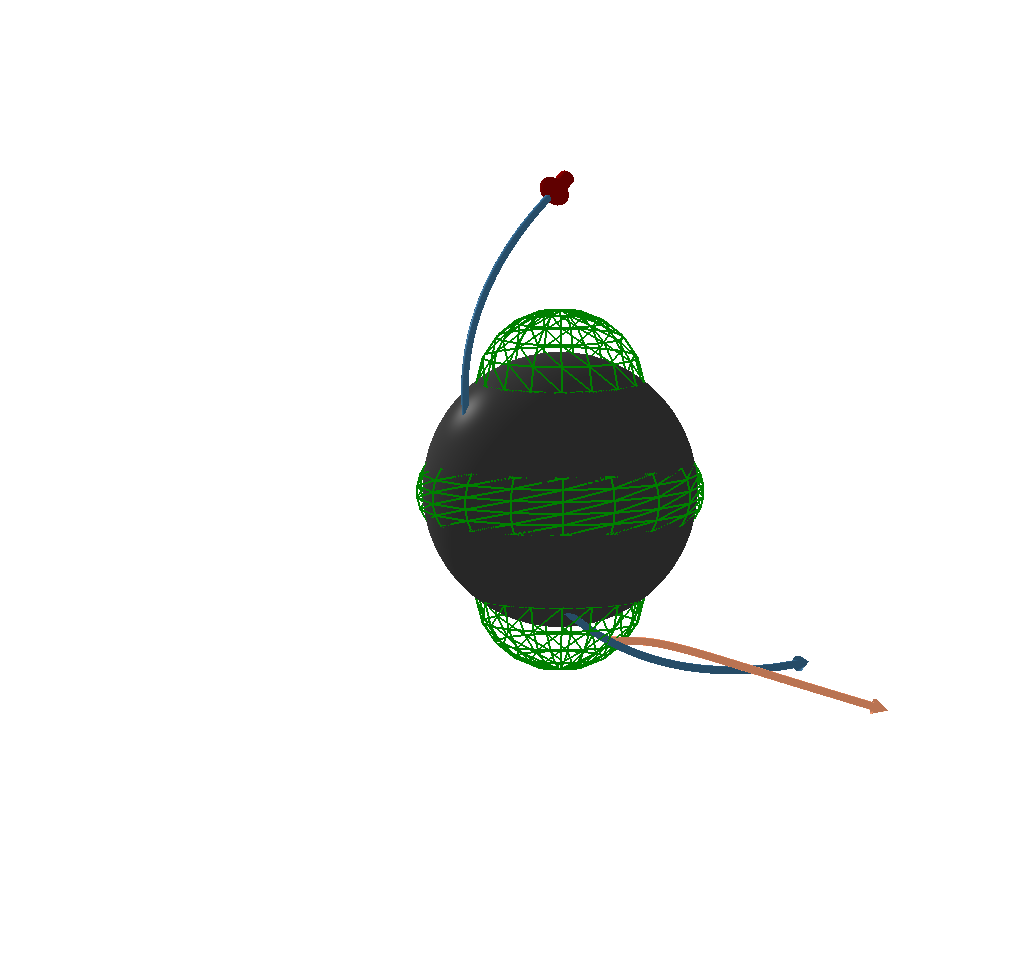}
        \includegraphics[width=.49\textwidth]{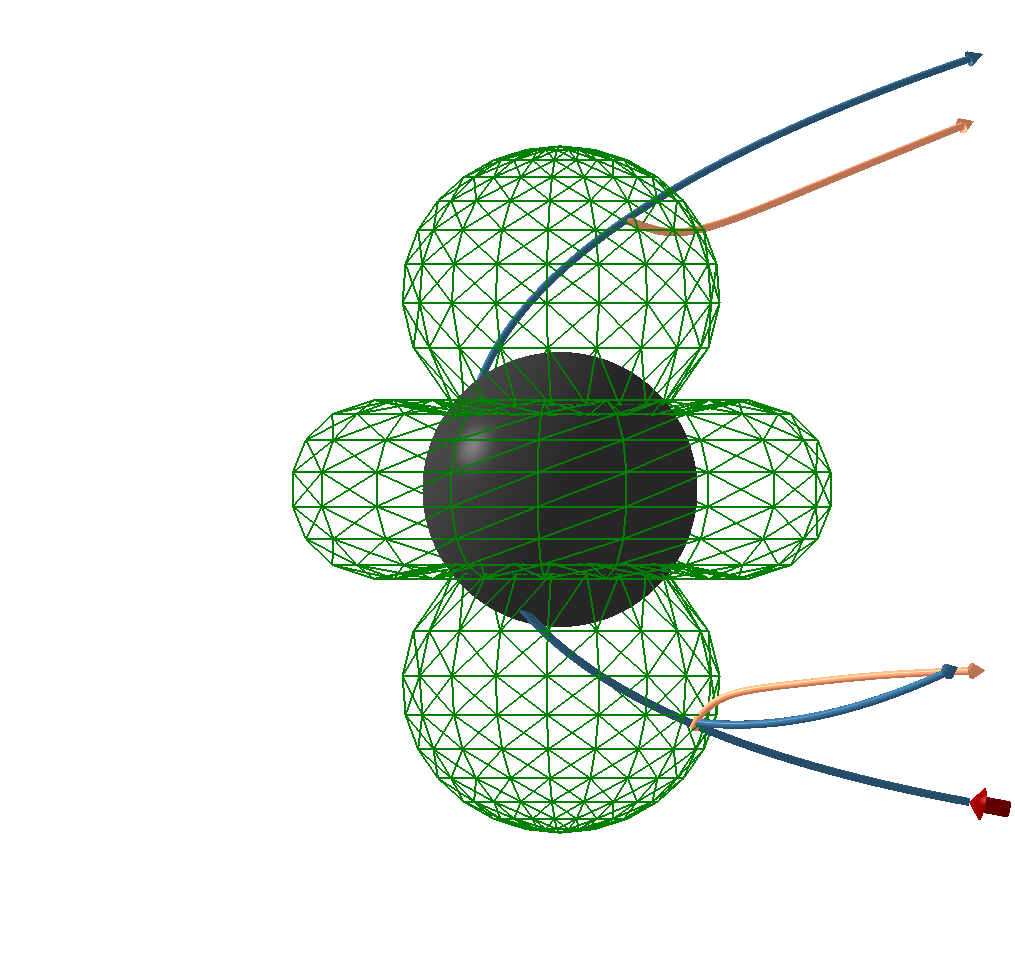}
        \includegraphics[width=.49\textwidth]{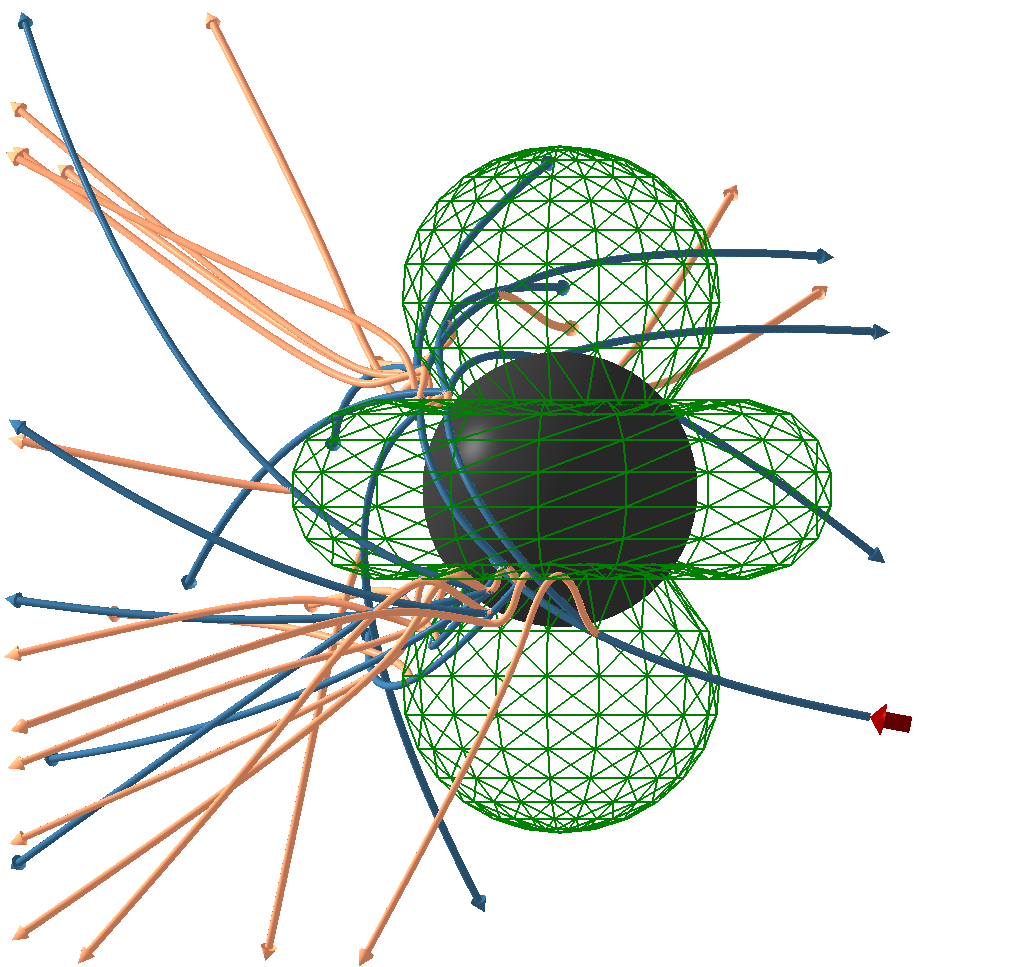}
    \caption{
    Illustration of possible final states  arising from a single infalling axion (initial trajectory marked with red arrow). The infalling axion encounters resonances when $\omega_p \simeq m_a$ (green surface), potentially converting to a photon (orange); the photon in turn may encounter resonances during its propagation, potentially reconverting back into  an axion (blue). This processes iterates until all possible outgoing trajectories have been identified. The collection of these final states, and their trajectories through the magnetosphere, are illustrated for two choices of axion mass (the upper panel illustrating the case of $m_a = 26 \, \mu$eV, and the lower panels illustrating the case of $m_a = 10 \, \mu$eV) and two differential sets of initial conditions. The top right panel illustrates the `single level crossing' scenario discussed in Sec.~\ref{sec:Results}, and the bottom right panel illustrates the potential complexity that can arise in these systems.
    }
    \label{fig:examples}
\end{figure*}

Axions falling through a neutron star magnetosphere can resonantly mix with low-energy electromagnetic modes when the four-momentum of the photon matches the four-momentum of the axion, \ie $k_\mu^\gamma = k_\mu^a$. In the highly magnetized plasma found in the inner magnetosphere, the only super-luminous electromagnetic mode that can be excited is the Langmuir-O (LO) mode, whose dispersion relation is given by~\cite{Witte:2021arp,Millar:2021gzs,McDonald:2023shx}\footnote{Note that corrections to the photon dispersion relation arising from the Cotton-Mouton term~\cite{Hochmuth:2007hk,Fairbairn:2009zi} and the Euler-Heisenberg term~\cite{Dobrynina:2014qba} are entirely negligible in these systems.}
\begin{equation}\label{eq:k_LO}
    \omega^2 = \frac{1}{2}\left(k^2 + \omega_p^2 + \sqrt{k^4 + \omega_p^4 + 2 \omega_p^2 k^2 (1-2\cos^2\theta_k)} \right) \, ,
\end{equation}
where $\omega_p = \sqrt{4\pi \alpha n_e / m_e}$ is the plasma frequency of the medium\footnote{Note that this expression is only valid for a non-relativistic plasma composed largely of $e^\pm$ pairs, see \eg\cite{Witte:2021arp,Millar:2021gzs}. These conditions are expected to hold along the closed field zones of pulsars, which comprise a majority of the region of interest.}, $k = |\vec{k}|$ is the modulus of the photon three-momentum and  $\theta_k$ is defined as the angle between the photon momentum and the magnetic field. Eq.~\eqref{eq:k_LO} can be used to solve for the location of the resonances, which occur when~\cite{Millar:2021gzs}
\begin{equation}\label{eq:res_p}
    \omega_p^2 = \frac{m_a^2}{m_a^2\cos^2\theta_k + 
    \omega^2 \sin^2\theta_k} \, ,
\end{equation}
which reduces in the non-relativistic limit to $\omega_p\simeq m_a$.
The efficiency of resonant LO mode production has been computed analytically using various approximation schemes (always assuming that the conversion takes place in a very narrow region near the resonance itself, an assumption which is expected to hold to high degree in most contexts)~\cite{Hook:2018iia,Witte:2021arp,Battye:2021xvt,Millar:2021gzs}, with the most recent calculation producing an axion-photon conversion probability given by~\cite{McDonald:2023ohd} 
\begin{equation}
     P_{a\to\gamma}^\text{non-ad} \simeq \frac{\pi}{2} \, \frac{g_{a\gamma}^2 \, |\vec{B}|^2 \, \omega_\gamma^4 \, \sin^2\theta_k}{\cos^2\theta_k \, \omega_p^2 \, (\omega_p^2 - 2 \omega^2) + \omega^4} \, \frac{1}{|\vec{v}_p \cdot \nabla_{\vec{x}} \omega|} \, , 
\end{equation}
where $\vec{v}_p = \vec{k}/\omega$ is the phase velocity of the photon at the resonance. This expression is only valid in the non-adiabatic limit ($P_{a\rightarrow \gamma} \ll 1$), but is expected to generalize in the adiabatic limit ($P_{a\rightarrow \gamma} \sim 1$) to the Landau-Zener formula~\cite{Battye:2019aco}\footnote{The Landau-Zener formula holds when the level crossing can be approximated as linear~\cite{Landau:1932vnv,Zener:1932ws} (see \eg\cite{Carenza:2023nck,Brahma:2023zcw} for examples of how the conversion probability change in more complex scenarios), which is thought to be a good approximation for axion-photon conversion near neutron stars.}
\begin{equation}\label{eq:LZ}
    P_{a\to\gamma}^{\rm ad} = 1 - \me^{ \gamma} \, ,
\end{equation}
where $\gamma = P_{a\to\gamma}^\text{non-ad}$ is the adiabaticity parameter.

Once produced, photons are refracted away from the neutron star by the dense plasma. Owing to the highly non-linear
trajectories of photons in this media, understanding the evolution and fate of the newly sourced photons requires dedicated ray tracing simulations (see \eg\cite{Witte:2021arp,Battye:2021xvt,McDonald:2023shx}), which amounts to solving Hamilton's equations, given by
\begin{eqnarray}
\frac{d x^\mu}{d\lambda} &=& \frac{\partial \mathcal{H}}{\partial k_\mu} \label{eq:hams1}\\
\frac{d k_\mu}{d\lambda} &=& -\frac{\partial \mathcal{H}}{\partial x^\mu} \, \label{eq:hams2},
\end{eqnarray}
where $\lambda$ is the wordline of the photon, and the photon Hamiltonian in a magnetized plasma is given by
\begin{eqnarray}
    \mathcal{H}(x^\mu, k^\mu) = g^{\mu\nu}k_\mu \, k_\nu + (\omega^2 - k_{||}^2) \, \frac{\omega_p^2}{\omega^2} \, .
    \label{eq:ham}
\end{eqnarray}
Here, we have introduced $k_{\|} = k \cdot B / \sqrt{B \cdot B}$, where the `$\cdot$' notation represents a contraction over the spatial indices, and the spatial dependence is understood to be implicitly embedded in $\omega_p$, the spacetime metric $g_{\mu \nu}$ and the magnetic 4-vector field $B_\mu$. Note that Eqns.~\eqref{eq:hams1}--\eqref{eq:hams2} can also be used to solve for axion trajectories, but using the simpler Hamiltonian given by $ \mathcal{H}_a(x^\mu, k^\mu) = g^{\mu\nu}k_\mu \, k_\nu - m_a^2$.

When computing the evolution of axion and photon trajectories, we use the Schwarzschild metric, taking a characteristic neutron star mass $M_{\rm NS} = 1 \, M_\odot$. For axions traversing the neutron star itself, we switch to the interior Schwarzschild metric~\cite{Schwarzschild:1916ae} (which assumes a constant density on the interior of the star), adopting in this case a neutron star radius $R_{\rm NS} = 10$ km. These values are merely intended as `ball-park' estimates, with measured systems suggesting typical neutron star masses closer to $M_{\rm NS} \sim 1.4 \, M_\odot$ and $R_{\rm NS} \sim 10-14$ km (see \eg~\cite{kiziltan2013neutron,lattimer2019neutron}); the impact of varying these parameters in the non-adiabatic limit has recently been discussed in~\cite{McDonald:2023shx}, and alternative choices are not expected to qualitatively alter any of the conclusions drawn based on the rough estimates used here.

The main purpose of this paper is to study the non-trivial evolution of the axion and photon phase space in the adiabatic limit. This is accomplished by: (1) following infalling axion dark matter particles as they fall through the magnetosphere, (2) identifying all resonance points $\mathcal{R}_i$ encountered during the infall, (3) assigning a phase space factor $\xi_i$ which accounts for the initial number density of infalling axions, the survival probability of the axion to reach resonance $\mathcal{R}_i$, and the probability of photon production (Eq.~\eqref{eq:LZ}), and (4) iteratively repeating steps (2--4) with the newly produced photons until all possible resonances have been identified and all axion and photon trajectories have been traced to asymptotic distances. Owning to the large number of resonances that can be encountered, this procedure can become quite complex -- as illustrated in Fig.~\ref{fig:examples}, a single infalling axion can lead to anywhere between $\mathcal{O}({\rm few})$ and $\mathcal{O}(10^3)$ possible outgoing axions and photons. The radio signal can be computed by summing over the asymptotic position of photon trajectories, localized in some region in the sky, where each photon trajectory is appropriately weighted by the initial axions phase space and the probability that it was produced and survived~\cite{McDonald:2023shx}. 

In order to make concrete quantitative statements about the behavior of the radio flux in the adiabatic regime we adopt a fiducial model for the magnetosphere characterized by a dipolar magnetic field and a fully charge-separated Goldreich-Julian (GJ) charge density, which can be derived by searching for the minimal co-rotation charge density necessary to screen $\vec{E}\cdot \vec{B}$ in the magnetosphere. The GJ charge density yields a plasma frequency near the neutron star of~\cite{Goldreich:1969sb}
\begin{eqnarray}
    \omega_p &\simeq& \sqrt{\frac{4\pi\alpha}{m_e} \frac{2 \,  \vec{\Omega} \cdot \vec{B}}{e}}  \nonumber \\
    &\simeq& \sqrt{\frac{e \, \Omega \, B_0}{m_e} \, \left( \frac{r_{\rm NS}}{r} \right)^3 \left|3\cos\theta \, \hat{m} \cdot \hat{r} - \cos\theta_m \right| } \, ,
\end{eqnarray}
where $\boldsymbol{\Omega}$ is the rotational frequency vector, $\hat{m}$ is the unit vector in the direction of the rotating magnetic dipole and $\theta_m$ is angle between the two. The factor $\hat{m} \cdot \hat{r} = \cos\theta_m \cos\theta + \sin\theta_m \sin\theta \cos(\Omega \, t)$ encodes the angular factor between the magnetic axis and the vector $\hat{r}$. The surface magnetic field strength is denoted $B_0$.

The fully charge-separated GJ model predicts small regions of vacuum, located at angles $\theta_{\rm null}$,  at the boundary of the charge separated regions\footnote{
    The approximate location of these regions of vacuum can be inferred from the bottom panels of Fig.~\ref{fig:examples}; they appear in the boundaries between the torus and dome-like features.
}.
While full-charge separation is expected to appear in dead neutron stars (see \eg~\cite{Spitkovsky:2002wg,Petri:2016tqe,Safdi:2018oeu}), it is unclear the extent to which these features survive for more active pulsars. As we will show in the following sections, the presence of vacuum regions extending to the neutron star surface can play an important role in controlling the efficiency of radio production at high couplings. In order to make conservative statements about the adiabatic regime, we thus also consider the inclusion of a small `boundary layer' of plasma around the neutron star, constructed in such a way that the large-scale features of the conversion surface (and plasma distribution) are unaltered, but the regions of vacuum near the neutron star are partially filled with a plasma density comparable to what is found at angles
$\theta \sim 0$ and $\theta = \pi/2$.
Specifically, in the case of an active pulsar, we consider an additive boundary layer contribution $\omega_{p, {\rm BL}}$ to the plasma density of the form
\begin{eqnarray}
    \omega_{p, {\rm BL}} = \omega_{p, 0} \,  \left(\frac{R_{\rm NS}}{r} \right)^{3/2} \, e^{- (r - r_b) / (\delta r)} \, ,
\end{eqnarray}
where $\omega_{p, 0}$ is the GJ plasma frequency at the pole, $r_b = 0.3 \times R_{\rm max}$ and $\delta r = 0.1 \times R_{\rm max}$, where $R_{\rm max}$ is the maximal radial extent of the conversion surface. The coefficients in $r_b$ and $\delta_r$ have been chosen in such a manner that the filling of the null lines is significant, but the plasma at $R_{\rm max}$ remains nearly unmodified. Other functional forms could also be adopted to perform the same function, however  the conclusions will not be significantly altered so long as the large scale features of the conversion surface remain unaltered.

In the following we perform our analysis using both the GJ model, and the GJ + boundary layer -- the former should be understood to be representative of a dead pulsar, and the latter as a conservative treatment of a more active pulsar\footnote{Technically, the GJ charge distribution is only expected to be representative in the closed magnetic field lines, however the open field lines are volumetrically tiny in the region of interest and are thus not expected to play any important role in the evolution of these systems.}.

\section{Tracing the phase space evolution}\label{sec:Trees}

Here, we extend the forward ray tracing algorithm developed in~\cite{Witte:2021arp,Foster:2022fxn,McDonald:2023shx} to self-consistently include the complete evolutionary tracks of all axions and photons sourced near the neutron star.
The details of this algorithm are outlined below.

\begin{figure}[t]
    \centering
    \includegraphics[width=\columnwidth, trim={0.0cm 0.0cm 0.0cm 0.0cm}, clip]{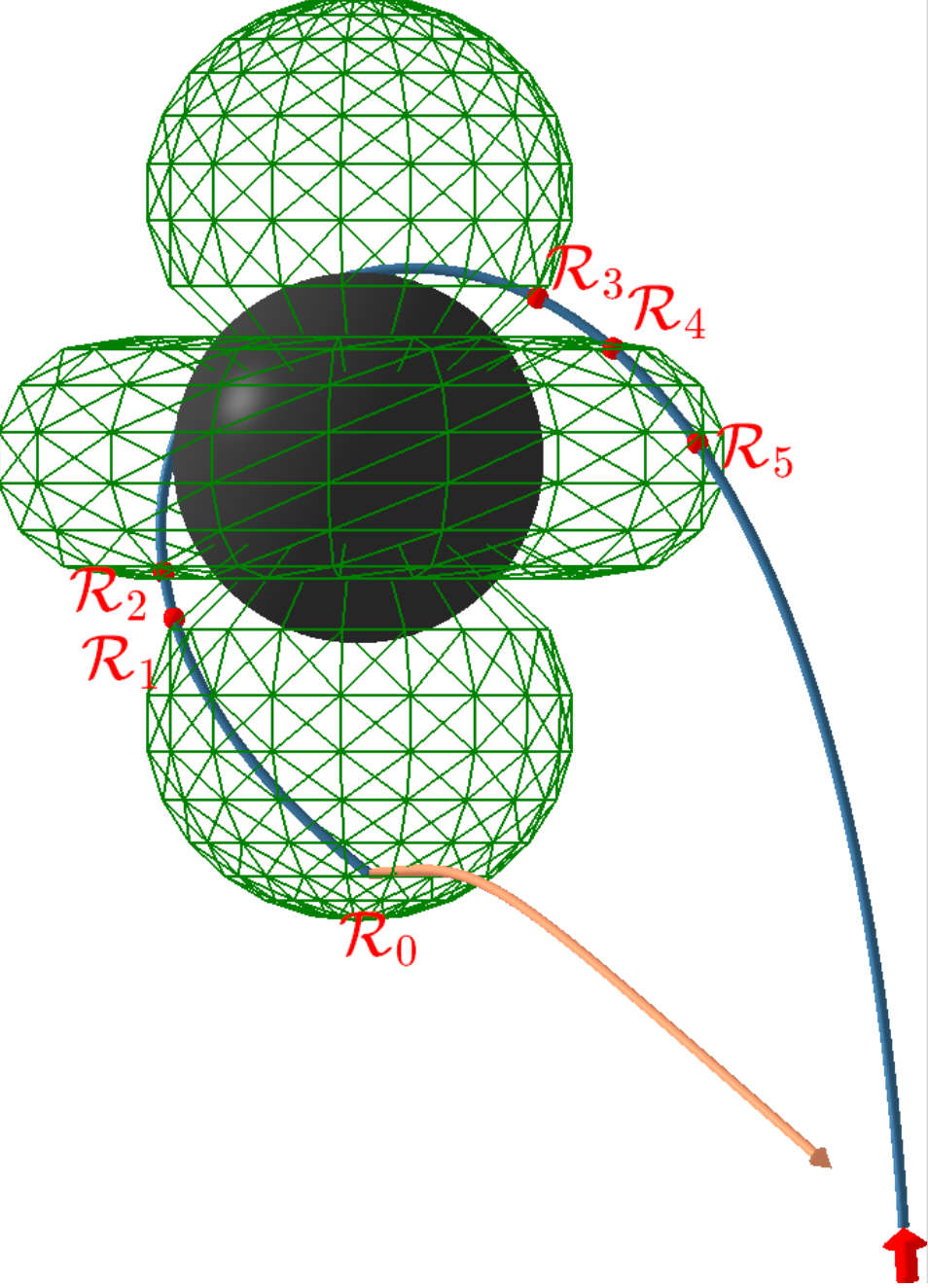}
    \caption{
    Similar to Fig.~\ref{fig:examples}, but highlighting instead the implementation of the numerical sampling procedure, which only traces subsets of the full trees illustrated in Fig.~\ref{fig:examples}. Here, an infalling axion is shown to cross several resonances (red points, labeled by $\mathcal{R}_{i>0}$) before converting to a photon at
    the MC sampled conversion point, $\mathcal{R}_0$. The final rate is re-weighted by the
    probability that the axion survives the resonances and converts into a photon at $\mathcal{R}_0$,
    $\xi_0^\gamma = 2n_{a,\infty}/\sqrt{\pi}\;P_{a\to\gamma}^0 \prod_{i=1}^5 P_{a\to a}^i$ (see Sec.~\ref{sec:RayTracing}).
    }
    \label{fig:procedure}
\end{figure}

We begin by applying the Monte Carlo (MC) surface sampling algorithm developed in~\cite{Witte:2021arp,McDonald:2023shx} to draw uniform samples from the resonant conversion surface, as defined in Eq.~\eqref{eq:res_p}. An axion trajectory is initially backward propagated away from this initial condition $\mathcal{R}_0=(\vec{x}_0, \vec{k}_0)$ to an asymptotic distance, and a phase space factor $\xi_i^a$ at each resonant point $\mathcal{R}_i$ is recorded\footnote{
    In practice, the sampling scheme applies Liouville's theorem
    to relate the phase space at infinity to the phase space
    on the conversion surface. This is, however, not applicable in
    the strong coupling limit. Therefore, the final weight of the
    event has to be re-weighted by the probability that the infalling
    axion indeed survives its travel to the sampled conversion points.
}, as indicated in Fig.~\ref{fig:procedure}. This factor accounts for the asymptotic axion energy density which sourced the initial infalling trajectory, the effect of gravitational focusing, and the probability that the infalling trajectory leads to a photon at $\mathcal{R}_0$.
In effect, this amounts to $\xi_0^a \equiv 2 n_{a,\infty} \, / \sqrt{\pi} \,(v_0 / v_\infty) \, P_{a\rightarrow \gamma}^{0}P_{a\rightarrow a}^{i>0} $, where $v_0$ is the axion velocity at $\xi_0$, $n_{a,\infty}$ and $v_\infty$ are respectively the asymptotic axion number density and velocity, and $P_{a\rightarrow a}^{i>0}$ represents the cumulative probability that the infalling axion is still an axion by the time it has reached the resonance $\mathcal{R}_0$\footnote{Note that this can be easily seen from the fact that the local axion density (under the assumption that the asymptotic axion distribution is homogeneous and isotropic) in the limit where $g_{a\rightarrow\gamma} \rightarrow 0$ is given by $n(r) \simeq 2 n_{a,\infty} \, / \sqrt{\pi} \,(v_0 / v_\infty) $~\cite{Hook:2018iia,Leroy:2019ghm}.}.
Starting from $\mathcal{R}_0$, we trace a photon trajectory out to infinity.
Since the photon may hit one or more conversion surfaces [see \eg Fig.~\ref{fig:examples}] -- potentially
re-converting into an axion -- all resonances encountered along its path must be tracked and assigned a weight $\xi_i^\gamma$. In turn, any newly sourced axions must also be propagated to infinity, and any conversion surfaces they
encounter be assigned a weight $\xi_i^a$.
This procedure is iterated until all possible resonances stemming for the original primary particle have been identified. Depending on the axion trajectory and the characteristic geometry of the conversion surface, each infalling axion trajectory can lead to anywhere from $\mathcal{O}({\rm few})$ to $\mathcal{O}(10^3)$ outgoing trajectories, making this a numerically challenging procedure.

The final radio flux at a given point on the sky $(\theta, \phi)$ can be obtained by taking the collection of outgoing photon trajectories which end up within a small angular bin on the sky (at asymptotic distances), and summing over the weighted contributions of each of these photons, \ie the power radiated in a region on the sky $(\theta_0 \pm \epsilon, \phi_0 \pm \epsilon)$ is given by
\begin{eqnarray}
    \mathcal{P}_{(\theta_0, \phi_0)} \simeq \frac{1}{N_s} \, \sum_i \mathcal{W}_i \, \mathcal{D}(\theta_{f,i}, \theta_0, \epsilon) \, \mathcal{D}(\phi_{f,i}, \phi_0, \epsilon) \, 
\end{eqnarray}
where $N_s$ are the number of samples drawn, 
\begin{eqnarray}
\mathcal{D}(x_i, x_0, \epsilon) = \begin{cases} 1 \hspace{.2cm} {\rm if} \, x_0 - \epsilon \leq x_i \leq x_0 + \epsilon \\
0 \hspace{.2cm} {\rm else} \, ,
\end{cases}
\end{eqnarray}
and we have defined the photon weight function $\mathcal{W}_i$, which in the sampling procedure of~\cite{Witte:2021arp,McDonald:2023shx} is given by\footnote{The pre-factors in Eq.~\eqref{eq:w_i} may differ depending on how one chooses to sample the phase space at the conversion surface; the result shown here is valid only for a uniform sampling procedure. }
\begin{eqnarray}\label{eq:w_i}
    \mathcal{W}_i \equiv N_{\rm max} \, (2 \pi {R}_{\rm max}^2)  \, |\cos\theta_{k \nabla E}| \, k \, \sqrt{|h_{\vec{k}}|} P_{a\rightarrow\gamma} \, n_{a, \rm loc}.
\end{eqnarray}
Here, we have defined $\cos\theta_{k \nabla E}$ as the angle between $\vec{k}$ and $\nabla E_\gamma$, the pull-back metric $h_{\vec{k}}$ which defines the momentum dependent conversion surface, the maximal number of resonant crossings per sample $N_{\rm max}$ and the maximal radial distance used in the surface sampling algorithm ${R}_{\rm max}$, and the local axion number density $n_{a, \rm loc}$.

\begin{figure*}
    \centering
    \includegraphics[width=\textwidth]{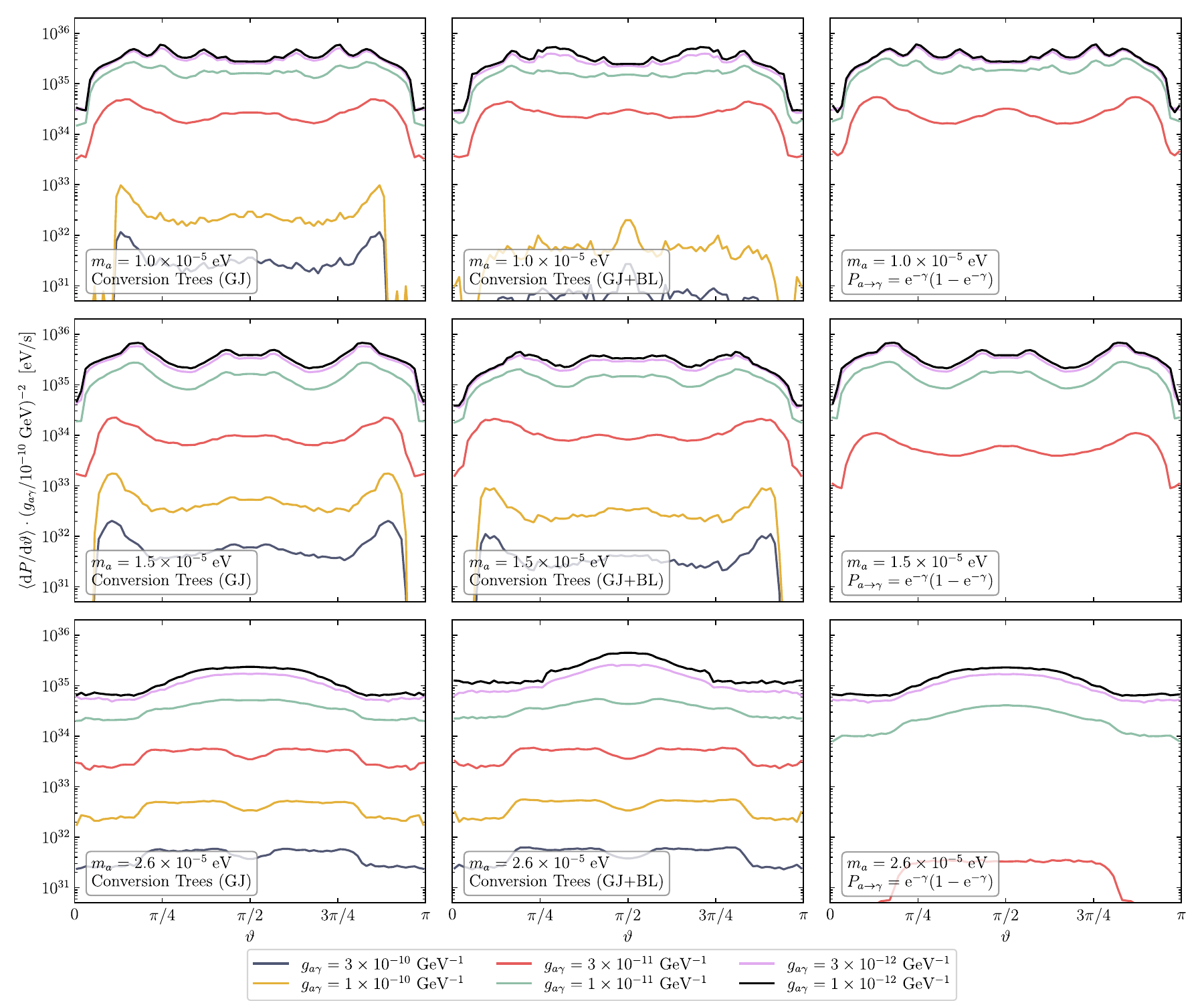}
    \caption{
    Differential radiated power from dark matter axions converting into
    photons in the magnetosphere of a neutron star for axion masses
    $m_a =1.0\times 10^{-5}\unit{eV}$ (first row),
    $m_a=1.5\times 10^{-5}\unit{eV}$ (middle row) and
    $m_a=2.6\times 10^{-5}\unit{eV}$ (bottom row), and for axion-photon
    couplings from
    $g_{a\gamma}=3\times 10^{-10}$ to $g_{a\gamma}=10^{-12}$ (see legend).
    The results are shown using the newly developed algorithm to compute the conversion trees from each infalling axion using either the GJ magnetosphere (left column), or the GJ magnetosphere with
    an additional boundary layer as described in Sec.~\ref{sec:RayTracing} (middle column).
    In addition, the results with the conservative
    approximation $P_{a\to\gamma}=\me^{-\gamma}(1-e^{-\gamma})$ are shown in the right
    column for comparison.
    }
    \label{fig:diff_theta}
\end{figure*}

\begin{figure*}
    \centering
    \includegraphics[width=\textwidth]{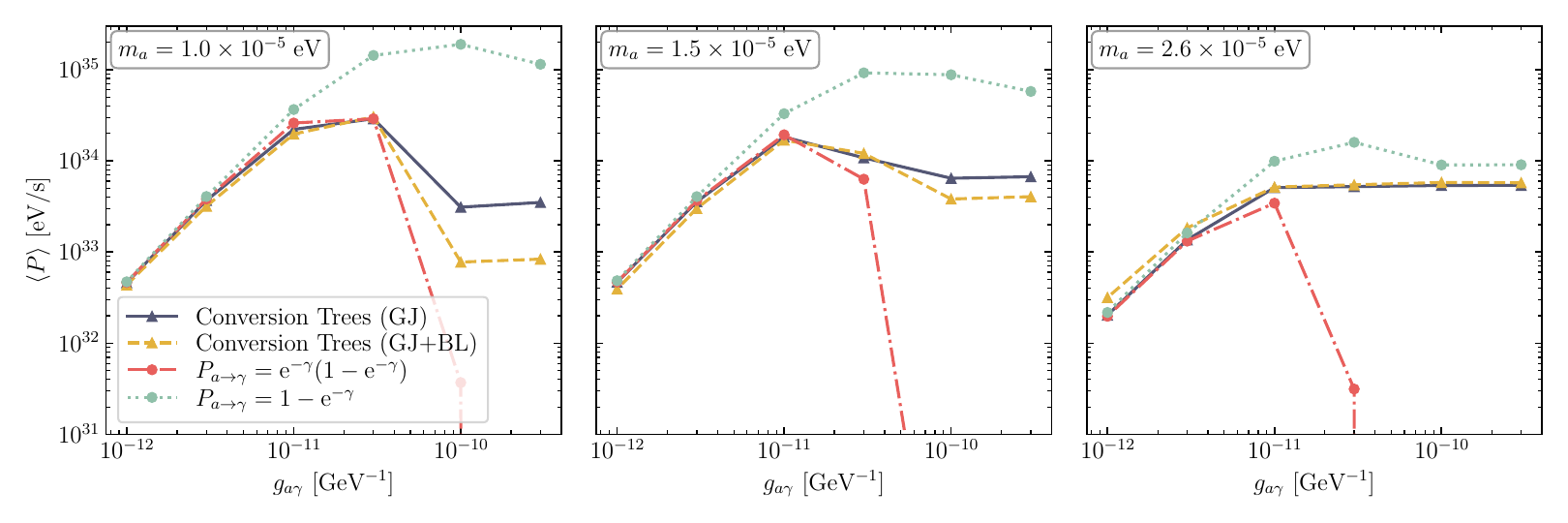}
    \caption{Integrated radiated power from dark matter axions converting into
    photons in the magnetosphere of a neutron star for axion masses
    $m_a =1.0\times 10^{-5}\unit{eV}$ (left plot),
    $m_a=1.5\times 10^{-5}\unit{eV}$ (middle) and
    $m_a=2.6\times 10^{-5}\unit{eV}$ (right).
    The results are shown for the conversion tree calculation with (blue solid line) and without (yellow dashed line) the boundary layer
    discussed in Sec.~\ref{sec:RayTracing}.
    For comparison, the results with the conservative
    approximation $P_{a\to\gamma}=\me^{-\gamma}(1-e^{-\gamma})$ are plotted (red dashed dotted line).
    In addition, we plot the results that would be obtained by assuming photons cannot re-covert to axions once produced -- this is obtained by taking
    $P_{a\to\gamma}=1-e^{-\gamma}$, and effectively sets an upper limit on the flux (green dotted line). The small reduction in the green line at high couplings arises from photons which impact the NS surface (and are thus lost).
    }
    \label{fig:total_flux}
\end{figure*}

The full tree, \ie all possible outcomes, of the path of the outgoing photon should be considered,
since even small axion-photon conversions may lead to detectable radio signals.
 The computational time for the full tree, however, is naively expected to scale as $\sim 2N_{\rm res}+1$, where $N_{\rm res}$ is the number of resonances encountered\footnote{Note that the number of out-going trajectories $N_{\rm out}$ scales with the generation $N_g$ as $N_{\rm out} = 2^{N_g}$, and the number of resonances scales like $N_{\rm res} = 2^{N_g - 1}$. These relations can be used to compute the total number of trajectories (including internal legs), which is given by $N_{\rm tot} = 1 + \sum_{i\leq N_g} N_{\rm out, i} = 1 + 2 \sum_{i\leq N_g} N_{\rm res, i} = 2  N_{\rm res} + 1$ (where the factor of one comes from the initial trajectory).}, making this procedure significantly  more computationally intensive than in the non-adiabatic limit\footnote{
    Note that the most computationally
    expensive part is the highly non-linear propagation near the
    resonances, implying that the computational time scales with
    the number of sub-branches in the tree, $2N_{\rm res}+1$.
}.
In order to avoid severe computational time
for complicated trajectories (see \eg bottom right panel of Fig.~\ref{fig:examples}), we
transition to a pure MC sampling after $N=5$ conversion points have
been encountered, implying that we consider (at most) 6 outgoing particles\footnote{
    In addition, we include two stopping criteria to hinder potential rare semi-stable
    or complicated trajectories. First,
    we truncate the MC simulation if 50 resonances are encountered; such events are rare, occurring in only 13 of the $\sim 10^7$ events included in the analysis. Second,
    the simulation is truncated when the simulated outcomes account for more than $1-10^{-100}$. This second threshold is overly conservative, and can be relaxed significantly to further reduce computation time.
}
-- thus making sure that we include at least one outgoing photon in the event.
In practice, this is achieved by always
considering the branch with the highest weight, \ie a particle is only propagated
until it reaches a resonance, the outcomes stored in a pool, and the particle in the
pool with the largest weight at any given time is propagated.

In most cases, the tree of possible outcomes is rather simplistic. For example,
the average number of resonances encountered, $N_\mathrm{res}$ considered in the next section are
3.1, 2.3, and 1.7, for the
masses $m_a=1.0\times 10^{-5}, 1.0\times 10^{-5}$ and $2.6\times 10^{-5}$ eV,
respectively, in the GJ magnetosphere\footnote{A number 1 means that only a single photon is
forward propagated and a single axion backwards propagated as in Fig.~\ref{fig:procedure}.
}.
Despite many trajectories being simple, the MC selection process was triggered (\ie more than 5 resonances encountered in 
the tree) in 13.3\%, 6.2\% and 0.1\% of the trees, contributing to as much as 26.0\%, 11.1\% and 0.1\% of the total flux at large couplings.

\section{Results}\label{sec:Results}
Using the algorithm discussed above, we analyze the radio signal emanating from a neutron star with a dipolar magnetic field with surface strength $B_0 = 10^{14}$ G, a rotational period $P = 2\pi {\rm s}$, and a misalignment angle $\theta_m$, which we set to zero for simplicity\footnote{We also take $M_{\rm NS} = 1 \, M_\odot$ and $r_{\rm NS} = 10$ km. The neutron star mass and radius are not expected to qualitatively change our conclusions (with the predominant effect being $\mathcal{O}(1)$ shifts in the overall flux~\cite{McDonald:2023shx}). Non-zero misalignment, on the other hand, has the dominant effect of smoothing out the differential power across a wider range of viewing angles (see \eg~\cite{Witte:2021arp}).  }. Owing to computational costs, we choose to keep these three parameters fixed throughout the analysis, varying only the axion mass, $m_a$, and axion-photon coupling, $g_{a\gamma}$. As we will show below, the scaling of the radio flux into the adiabatic regime is largely set by the geometry of the conversion surface -- since shifting $B_0$ and $P$ alter the geometry in a manner that is fully degenerate with a shift in the axion mass, and the role of $\theta_m$ is at leading order to induce a small rotation in the conversion surface, we believe the results identified here are quite general\footnote{The only notable subtlety is that the axion-photon coupling at which the adiabatic regime is encountered can shift to smaller or larger values, depending on the magnetic field strength and the plasma density in the magnetosphere.  For this reason, our results should be interpreted qualitatively rather than quantitatively.  }.

In Fig.~\ref{fig:diff_theta}, we plot the period-averaged differential power as viewed from an angle $\theta$ with respect to the axis of rotation, for three different axion masses and six choices of the axion-photon coupling (which smoothly extend the results from the non-adiabatic to adiabatic regimes). Each of the differential power curves have been re-scaled by a factor of $(g_{a\gamma} / 10^{-10} \, {\rm GeV^{-1}})^{-2}$, so that the suppression of the power in the adiabatic regime can be more easily identified (in the non-adiabatic regime, this re-scaling causes all curves to lie on top of one another).  We illustrate the evolution of the differential flux at large couplings using three distinct approaches. In the right column, we adopt the approximation scheme of~\cite{Foster:2022fxn}, which amounts to assigning each photon an effective conversion probability $P_{a\rightarrow \gamma, {\rm eff}} = \me^{-\gamma}(1-\me^{-\gamma})$. The left column, instead, shows a comparison with the full conversion tree as computed using the algorithm described in the preceding section. Finally, in the center column, we compute the full conversion tree including the `boundary layer' contribution to the plasma frequency discussed in Sec.~\ref{sec:RayTracing}.

\begin{figure}
\centering
\includegraphics[width=\columnwidth]{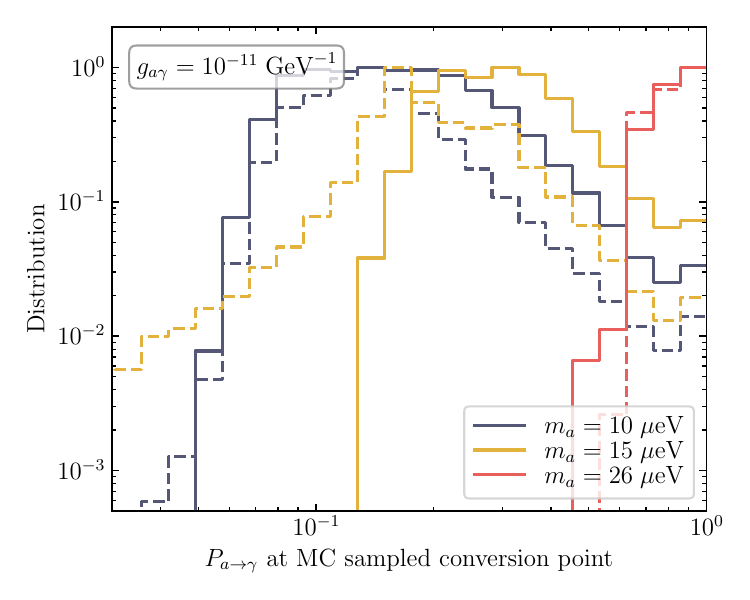}
\caption{Distribution of the axion-photon conversion probability at
    the MC sample points $\vec{x}_0$.
    The distribution is shown for the three masses
    $m_a=1.0\times 10^{-5}\unit{eV}$ (blue),
    $m_a=1.5\times 10^{-5}\unit{eV}$ (yellow), and
    $m_a=2.6\times 10^{-5}\unit{eV}$ (red).
    The solid lines indicate the GJ model, while the dashed includes the additional boundary
    layer.
    }
\label{fig:hist}
\end{figure}

A number of features can be readily appreciated from Fig.~\ref{fig:diff_theta}. First, the naive approximation scheme of ~\cite{Foster:2022fxn} tends to consistently over suppress the flux in the adiabatic regime. Next, the suppression is largely, but not entirely, uniform across the sky -- for small axion masses, the suppression is more apparent near the magnetic poles, but away from the poles the suppression is more uniform (note that in the case of misaligned rotators, this effect would be smeared across viewing angles). In addition, the suppression appears to be much more prominent for small axion masses, which corresponds to the scenario where the resonant conversion surface extends further from the neutron star surface (see Fig.~\ref{fig:examples}). Finally, the existence of a small boundary layer of plasma around the neutron star tends to suppress the flux relative to the GJ magnetosphere, but not as much as the effective scheme adopted in ~\cite{Foster:2022fxn}. 

These features can also be appreciated by looking at the sky-averaged flux as a function of the axion-photon coupling; Fig.~\ref{fig:total_flux} compares each of the three models for all three axion masses. Here, one can see that the radio flux is actually expected to plateau at sufficiently large axion-photon couplings, rather than become exponentially suppressed. The relative height of the plateau depends both on the axion mass and the on the existence of charge-separation in the magnetosphere.

Collectively, Figs.~\ref{fig:diff_theta} and~\ref{fig:total_flux} lead to two significant conclusions:
\begin{itemize}
\item Despite the fact particles, on average, encounter an even number of level crossings, the efficiency of these level crossings is not equivalent. As such, the approach of  ~\cite{Foster:2022fxn} naturally over-estimates the suppression of the radio flux in the adiabatic regime.
\item Ref. ~\cite{Foster:2022fxn} missed the importance of infalling axion trajectories which only encounter a single resonance, which despite often being uncommon can dominate the radio flux. Such trajectories occur when axions traverse the neutron star, entering or exiting near the charge separation boundary. For large axion masses, a larger fraction of the neutron star surface is `exposed', allowing for a larger fraction of the infalling axion phase space to encounter single level crossings. 
\end{itemize}

\begin{figure}
\centering
\includegraphics[width=\columnwidth]{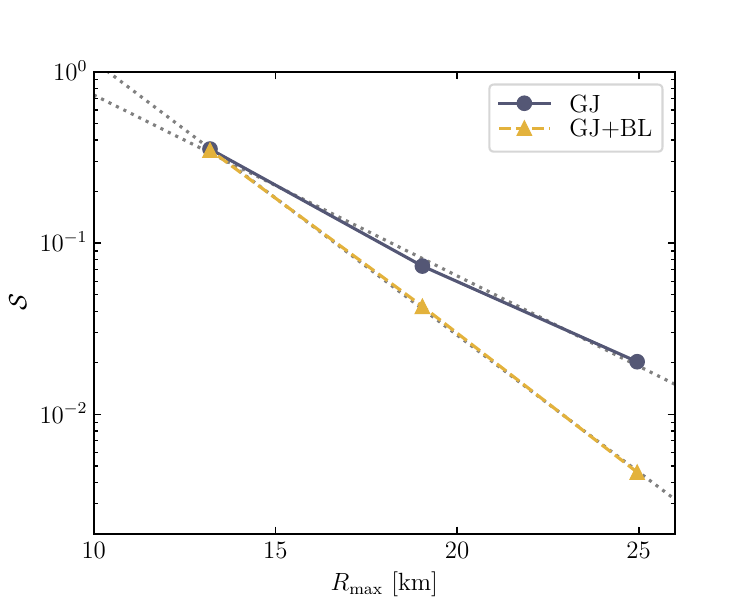}
\caption{
Suppression factor at large couplings as a function of $R_\mathrm{max}$ for the GJ
magnetosphere (blue solid line) and including the additional boundary layer (yellow
dashed line). The gray dotted lines indicate the fitted functions
$\mathcal{S}(R_\mathrm{max}) = \exp(-0.243 R_\mathrm{max} + 2.124)$ [GJ] and
$\mathcal{S}(R_\mathrm{max}) = \exp(-0.369 R_\mathrm{max} + 3.834)$ [GJ+BL].
}
\label{fig:suppression}
\end{figure}

\subsection{On the application to future searches}

The computational cost of running the full conversion tree makes it difficult to fully embed within a more sophisticated analysis of radio data, such as the analysis performed in ~\cite{Foster:2022fxn}. As such, we attempt to develop below an approximate re-scaling technique that can be adopted in future work to approximate the suppression in the flux that  arises in the adiabatic regime. 

\begin{figure*}
\centering
\includegraphics[width=\textwidth]{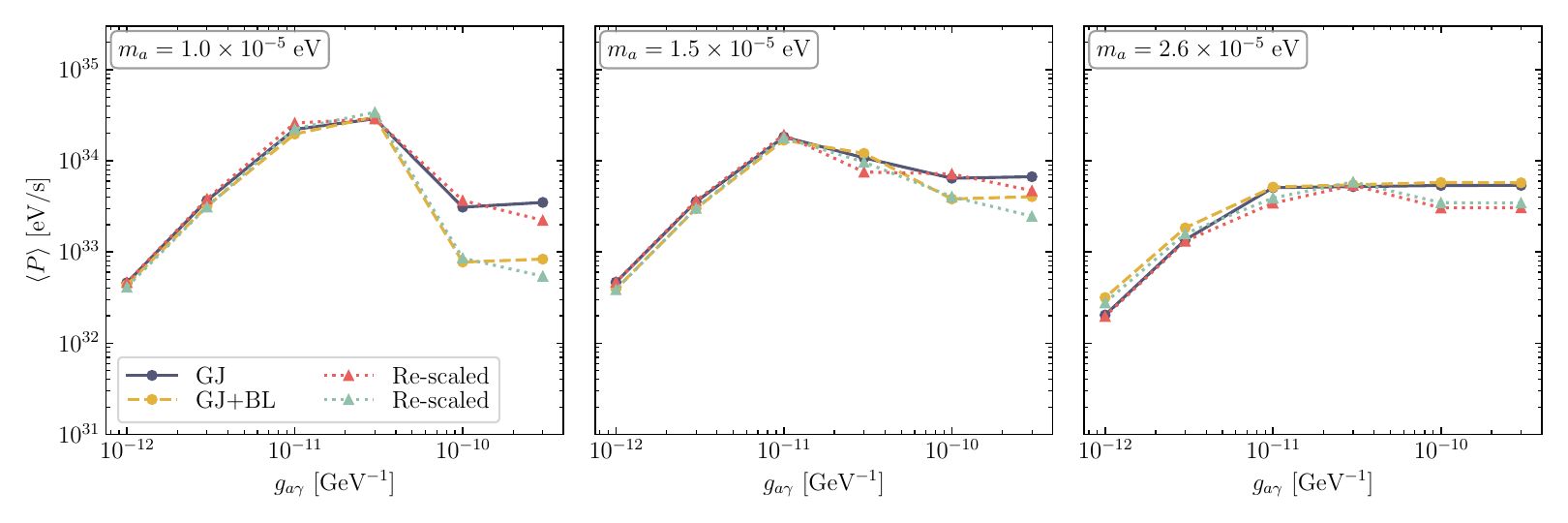}
\caption{
Same as Fig.~\ref{fig:total_flux}, but including the re-scalings discussed
in the main text.
}
\label{fig:reweight}
\end{figure*}

The simplest approximation that one can make to account for the transition from the non-adiabatic to the adiabatic regime, is that of~\cite{Foster:2022fxn}, $P_{a\to \gamma} = \me^{-\gamma} (1-\me^{-\gamma})$. However, as we have shown, this approach is overly conservative in the adiabatic limit (\cf Fig.~\ref{fig:total_flux}), and can lead to a significant underestimation of the total radio flux. On the other hand, the most optimistic approximation one can make is $P_{a\to \gamma}=1-\me^{-\gamma}$, \ie only a single level crossing.

An alternative approach is to try and encode the suppression of the radio flux of each neutron star into an effective re-scaling parameter, which depends\footnote{
The global suppression factor reflects the overall suppression for most viewing angles, with the exception near the poles, as can be deduced from Fig.~\ref{fig:diff_theta}. However, for slight misalignment angles, the in-homogeneity of the suppression is expected to be largely washed out.
} on $R_{\rm max}$. These suppression factors can then be included as `effective' conversion probabilities which are turned on as the neutron star enter the adiabatic regime,  $P_{a\rightarrow \gamma} \gtrsim 0.2$ [\cf Fig.~\ref{fig:hist}]. Using the  data points in our sample we derive approximate re-scaling factors $\mathcal{S}=\Phi(g_{a\gamma}\to \infty)/\Phi(P_{a\to\gamma}=1)$ for the sky-integrated flux -- the suppression factors are shown as a function of $R_{\rm max}$ in Fig.~\ref{fig:suppression}. We implement these suppression factors by adopting the adiabatic approximation at low couplings, $P_{a\to \gamma} = \me^{-\gamma} (1-\me^{-\gamma})$, and transitioning to our effective re-scaling  $P_{a\to\gamma}=\mathcal{S}(R_{\rm max}) \times (1-\me^{-\gamma})$ at larger values of $g_{a\gamma}$ -- in the intermediate regime, we adopt the maximum of the two approaches. We illustrate the relative agreement of applying this approach to the sky-integrated flux in Fig.~\ref{fig:reweight}.

\section{Conclusion}\label{sec:conclusions}

In this work, we have for the first time identified the behavior and scaling of radio emission produced from the resonant conversion of axion dark matter near neutron stars in the adiabatic (\ie strong mixing) limit. We have done this by developing an MC sampling and ray tracing algorithm capable of carefully tracking the evolution of the axion and photon phase space. 

Our results clearly indicate: (i) contrary to previous approximations, the radio flux is not exponentially suppressed at large axion-photon couplings, but rather plateaus to a fixed value, and (ii)
the radio flux is not suppressed at all for small conversion surfaces (\ie  $R_\mathrm{max}\sim R_\mathrm{NS}$), which would be the case if there exist dead neutron stars in the field of view which support a maximal plasma density only slightly in excess of the axion mass. We further illustrate an approximate scaling relation which can be used to extrapolate radio observations into the adiabatic regime, circumventing the need to apply computationally expensive simulations -- such as the one developed here -- to large numbers of systems. 

Our conclusions are based on a number of assumptions, most of which are thought to be well-justified; for the sake of clarity, we enumerate these assumptions below:
\begin{enumerate}
    \item Axion-photon transitions are dominated by the resonant contribution, and it is valid to treat the resonance with the WKB approximation (\ie that the  background varies slowly relative to the axion wavelength). For axion masses capable of generating radio emission, this approximation is expected to be true over a majority of the magnetosphere, with the one exception perhaps being regions near the return currents and open field lines.
    \item The adiabatic generalization of the non-adiabatic conversion probability is assumed to follow the Landau-Zener formula. This has been shown to be true in one-dimension and in an isotropic plasma (at least when the medium is smoothly and slowly varying) (see e.g.~\cite{Battye:2019aco}), but has not be explicitly derived for a three dimensional anisotropic plasma. 
    \item Axions are assumed to be fully non-interacting away from the resonance, and the axion population is assumed to arise exclusively from either in-falling axion dark matter, or from axions sourced from photons which themselves were sourced from axion dark matter (that is to say, local radiation from the magnetosphere is neglected). 
    \item No other exotic particle content is assumed to exist which could either alter the dispersion relations of these particles, or the assumption that their interactions with the ambient medium can be neglected.
    \item The magnetosphere is assumed to be approximately characterized by a purely dipolar field and the Goldreich-Julian charge density. Higher-order magnetic multi-poles may exist near the star, but are not expected to have a large qualitatively impact on our results. For standard pulsars, deviations from the GJ model are expected along the open field lines and near the return currents, but the closed field lines (comprising nearly all of the near-field magnetosphere) are roughly expected to be well-characterized by the GJ values. The charge distribution may differ notably for millisecond pulsars, magnetars, and binary pulsar systems; however, given a model of any of these systems, the formalism developed here can be applied to make quantitative statements about each of these.
\end{enumerate}

This work has important implications for the radio searches for axion dark matter (such as those performed in~\cite{Foster:2020pgt,Foster:2022fxn,Battye:2021yue}), and will prove important as these observations are extended to other systems and to a broader range of frequencies.

\section*{Acknowledgments}%
The authors thank Manuel Linares for catching a small typo. SJW acknowledges support from the Royal Society University Research Fellowship (URF-R1-231065), and through the program
Ram\'{o}n y Cajal (RYC2021-030893-I) of the Spanish
Ministry of Science and Innovation.
JT would like to express gratitude for the hospitality at the University of Agder (UiA). This article/publication is based upon work from COST Action COSMIC WISPers CA21106, supported by COST (European Cooperation in Science and Technology). JIM is grateful for the support of an FSR Fellowship and funding from the Science and Technology
Facilities Council (STFC) [Grant Nos.~ST/T001038/1 and ST/X00077X/1].

\bibliography{references}

\end{document}